\documentclass{aa}
\usepackage{graphicx}
\begin{document}
   \title{Low-extinction windows in the inner Galactic Bulge}

   \author{C.M. Dutra\inst{1,2}, B.X. Santiago\inst{1} \and E. Bica\inst{1}}

   \offprints{B. X. Santiago  -- santiago@if.ufrgs.br}

\institute{Universidade Federal do Rio Grande do Sul, IF, 
CP\,15051, Porto Alegre 91501--970, RS, Brazil\\
 \and
 Instituto Astronomico e Geofisico da USP, CP\, 3386, S\~ao Paulo 01060-970, SP,
 Brazil\\
}

   \date{Received ; accepted }

\abstract{
We built {\bf K band extinction maps in the area} of two candidate 
low-extinction windows in the inner  Bulge: W0.2-2.1 at 
{\bf ($\ell$,{\it b}) = (0.25$^{\circ}$,-2.15$^{\circ}$)}, and W359.4-3.1 at 
{\bf ($\ell$,{\it b}) = (359.40$^{\circ}$,-3.10$^{\circ}$)}. {\bf We employed  
{\it JHK$_s$} photometry from the 2MASS Point Source Catalog.} 
Extinction values were determined by fitting the upper giant branch found 
in the present 2MASS {\it $K_s$ ($J-K_s$)} diagrams to {\bf a de-reddened 
bulge stellar population reference giant branch.}  We tested  
the method {\bf on the well known Baade's and Sgr I windows: the 2MASS 
mean extinction values in these fields agreed well with those of 
previous studies. 
We confirm the existence of low-extinction windows in the regions studied,
as local minima in the $A_K$ maps reaching $A_K$ values about 2 standard deviations below
the mean values found in the neighbouring areas. Schlegel et al.'s (1998) FIR extinction maps, which integrate dust 
contributions throughout the Galaxy, are structurally  similar to 
those derived with 2MASS
photometry in the two studied windows. We thus conclude that  
the dust clouds affecting the 2MASS and FIR maps in these directions 
are basically the same and are located   
on foreground of the bulk of bulge stars. However, the {\it $A_K$} 
absolute values differ significantly. In particular, the
FIR extinction values for W359.4-3.1 are a factor $\simeq 1.45$ larger 
than those derived from 
the 2MASS photometry. Possible explanations of this effect are discussed.
The lower Galactic latitudes of the low-extinction 
windows W359.4-3.1 and W0.2-2.1, as compared to Baade's Window, make
them promising targets for detailed studies of more central bulge regions.}
\keywords{The Galaxy: interstellar medium: dust}}

\titlerunning{Low-extinction windows in the inner Galactic Bulge}
\authorrunning{C. M. Dutra et al.}

\maketitle

%
 
\section{Introduction}

Most of the Bulge stellar population is still largely unstudied due to the
combined effects of large distances and {\bf high extinction}.
The information on Bulge {\bf stellar} populations comes mainly from either globular clusters
or from its M and K field giant stars. Yet, the study of the age and metallicity distribution in 
this region is of considerable importance, since a comprehensive study of the resolved Galactic 
Bulge allows us to better understand the bulge of external early and late type galaxies.
Furthermore, inferring the main properties of the Bulge and comparing 
them to those of
other components of the Galaxy is likely to provide clues to unveiling the
process of galaxy formation (Aguerri et al. 2001, Wyse et al. 2000).

During the last years the Galactic Bulge stellar population has been studied {\bf mainly in the direction of low-extinction 
regions} (Lloyd Evans 1976, 
Whitford 1978, Terndrup {\bf 1988}, Tiede et al. 1995, {\bf Alard et al. 2001}). Baade (1963) identified {\bf the windows Sgr I,  Sgr II, and the 
NGC6522 field; this latter has been widely referred to as Baade's Window in subsequent studies. Stanek (1996) studied the extinction distribution in Baade's Window} using the OGLE photometry of red clump stars, obtaining values from 
{\it $A_V$} = 1.26 up to {\it $A_V$} = 2.79. Frogel et al. (1999, hereafter FTK99) determined the extinction 
for {\bf 11} inner Bulge fields using {\bf Baade's Window  red giant branch} as a reference, 
yielding values in the range  {\it $A_V$} = 2.41 up to 19.20. 

Recently, {\bf wide-angle} near infrared (NIR) surveys such as the 
Two Micron All Sky Survey (hereafter, 2MASS; 
Skrutskie et al. 1997) and the Deep NIR Southern Sky Survey (DENIS; 
Epchtein et al. 1997) have 
allowed investigations of the stellar population 
(Unavane et al. 1998) and reddening (Schultheis et al. 1999) in 
the inner Bulge. Schultheis et al. (1999) mapped the extinction in the inner 
Bulge for $|\ell|$ $<$ 8$^{\circ}$  and {\it $|b|$} $<$ 1.5$^{\circ}$ 
(with a resolution of {\bf 4$^{\prime}$}) using isochrone fitting to the 
colour-magnitude diagrams (CMDs) obtained from DENIS 
{\it J}, {\it K$_S$} observations. The extinction varies from 
{\it $A_V$} $\approx$ 6 up to {\it $A_V$} $\approx$ 37. 
They showed that the extinction and, as a consequence, the dust clouds 
in the inner Bulge, present a very patchy distribution.

In the present study we use the 2MASS survey in the {\it J} (1.25$\mu$m), 
{\it H} (1.65$\mu$m) and {\it K$_s$} 
(2.17$\mu$m) bands to {\bf identify} low-extinction windows in the inner 
Galactic Bulge. In  Sect. 2 
we discuss the process to select low-extinction candidate regions in the 
inner Galactic Bulge. In Sect. 3 we discuss
the method of deriving extinction values throughout these regions, 
provide a reddening 
distribution map for them and analyse the results.  Finally, the concluding 
remarks are given in Sect. 4.

\section{Selecting low-extinction windows}

 In search for candidate low-extinction regions we
used the DIRBE/IRAS dust emission redddening map from Schlegel et al. 
(1998, hereafter SFD98), which is available 
in Web Interface {\it http://astro.berkeley.edu/dust}. {\bf The extinction 
maps derived from these far infra-red
observations will be referred to as ``FIR extinction maps`` throughout 
the present work. Analogously, we will use the notation $A_{K,FIR}$ to
denote the K band extinction derived from SFD98 data. 
Stanek (1998) identified} two 
low-extinction windows 
{\bf on the FIR extinction maps, centred at ($\ell$,{\it b}) = (0$^{\circ}$,-2$^{\circ}$) and ($\ell$,{\it b}) = (4$^{\circ}$,-3$^{\circ}$).}
Although the {\bf FIR extinction} corresponds to the contribution of the 
entire dust column 
throughout the {\bf Galaxy} (Dutra \& Bica 2000), {\bf its angular distribution 
can help with the selection of potentially interesting}  areas of 
low-extinction towards the Bulge. We {\bf obtained} {\it E(B-V)$_{FIR}$} values 
in the field of 10$^{\circ}$$\times$10$^{\circ}$ around the Galactic 
Centre from the original {\bf overall FIR map in SFD98}, using {\bf an 
extraction tool thereby provided}. Figure 1 shows the resulting 
{\bf {\it $A_{K,FIR}$} extinction map, where the positions of} the 
three low-reddening regions {\bf from Baade (1963), two from Stanek (1998), and  
a newly identified one, are indicated.
The transformation from {\it E(B-V)$_{FIR}$} to {\it $A_{K,FIR}$} 
assumed {\it $A_K$} = 0.112 {\it $A_V$} and 
{\it R$_V$} = {\it A$_V$} / {\it E(B-V)} = 3.1 (Cardelli et al. 1989).} 

{\bf  In Table 1 we condense the basic information on the mentioned 
bulge windows. By columns: (1) adopted designation throughout the present study
 (a practical way to refer to the windows is ``W`` followed by its values of
  galactic coordinates), (2) other designations, (3) and (4) galactic
   coordinates, (5) approximate angular dimensions, (6) reference for
    window identification, and  (7) extinction map and photometric
  source.}
  
\begin{table*}
\caption[]{Low-extinction windows towards central parts of the Galaxy}
\begin{scriptsize}
\label{tab1}
\renewcommand{\tabcolsep}{0.9mm}
\begin{tabular}{lcccccc}
\hline\hline
Name & Other Designation  & $\ell(^{\circ}$)&    b($^{\circ}$)& d($^{\prime}$) & Reference & Photometry \\
\hline
W359.4-3.1 & -------- & 359.40&  -3.10& 40 $\times$ 30 &  present study & 2MASS  (present study)\\
W0.2-2.1 & (l,b)=(0,-2) Window &      0.24&  -2.14 &  60 $\times$ 40&    Stanek(1998)& 2MASS  (present study) \\
Baade's Window & NGC6522 field,W1.0-3.9 & 1.04& -3.88 &  60 &        Baade(1963)&           OGLE  (Stanek 1996)\\
Sgr I  Window&   W1.4-2.6 &  1.44& -2.64 &  60 &   Baade (1963)& ---------\\
W4.0+3.0 &   (l,b)=(4,3) Window  &    4.00&  3.00 &  180$\times$ 100 &Stanek(1998)&---------\\
Sgr II  Window  & W4.2-5.1 & 4.15 & -5.14  &    85  & Baade (1963) & ---------\\
\hline
\end{tabular}
\end{scriptsize}
\end{table*}

\begin{figure}
\centering
\resizebox{\hsize}{!}
 {\includegraphics{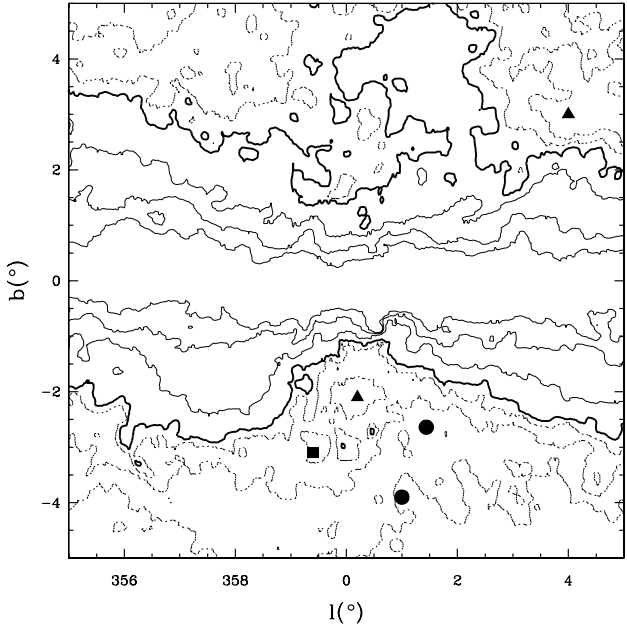}}
\caption[]{{\bf 10$^{\circ}\times 10^{\circ}$ FIR extinction map centred on 
the Galactic Nucleus. 
The line width is coded according to the $A_{K,FIR}$
value. Dotted lines correspond to lower than average levels: 0.7, 1.05 and 
1.4. The thick solid line is close to the field average and corresponds to
$A_{K,FIR} = 1.75$. The thin solid lines are above average: 3, 6, 10.
Filled circles indicate positions of windows from Baade (1963), triangles 
from Stanek (1998), and the square
from the present study.}}
\label{fig1}
\end{figure}

\section{Extinction within the candidate windows}
 
\subsection {FIR extinction maps}

Figure 2 shows the {\it $A_{K,FIR}$} maps within 2$^{\circ}$ $\times$ 
2$^{\circ}$ around {\bf ($\ell$,{\it b}) = (0$^{\circ}$, -2$^{\circ}$).
This region should contain W0.2-2.1, listed in Table 1 and which was 
originally identified by Stanek (1998). 
In fact, we can more precisely locate this window at 
($\ell$,{\it b}) = (0.25$^{\circ}$,-2.15$^{\circ}$), with 
dimensions $\approx$ $60^{\prime}\times 40^{\prime}$. This is a 
nearly closed region in Figure 2 where $A_{K,FIR}$ values 
are systematically below 
the average value in the map, {\it $<A_{K,FIR}>$} = 0.41. 
The lowest $A_{K,FIR}$
values within W0.2-2.1 reach down to $A_{K,FIR} = 0.28$, which corresponds
to about 2 standard deviations (std) below the mean. Therefore, there is 
a well defined local $A_{K,FIR}$ minimum in this area.}

Figure 3 shows a similar map {\bf centred at ($\ell$,{\it b}) = (0$^{\circ}$, -2$^{\circ}$),
close to the new candidate low-extinction region proposed in this work, 
W359.4-3.1. 
A distinct region with lower than average $A_{K,FIR}$ values is again visible;
it is centred at ($\ell$,{\it b}) = (359.40$^{\circ}$,-3.10$^{\circ}$), with 
$40^{\prime}\times 30^{\prime}$ in size. This region has a deeper minimum,
reaching $A_{K,FIR} = 0.24$, which is 2.5 std below the average map value
of {\it $<A_{K,FIR}>$} = 0.42. It is also more circular in shape and completely
encircled by areas of larger $A_{K,FIR}$ values.
We should point out that, by inspection of the FIR extincion map in Figure 1, 
one notices that even the average $A_{K,FIR}$ values in Figures 2 and 3 are
very atypical for their ($\ell$,{\it b}) location.}

\begin{figure}
\centering
\resizebox{\hsize}{!}{\includegraphics{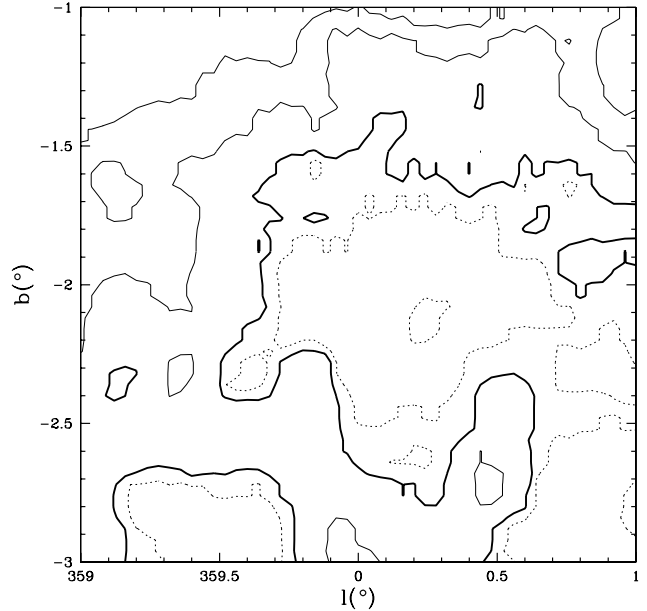}}
\caption[]{{\bf FIR  extinction map of window W0.2-2.1. 
Contour levels correspond to {\it $A_{K,FIR}$} =  0.29 and 0.37 (dotted lines),
0.41 (thick solid line), and  0.57 and 0.83 (thin solid lines).}}
\label{fig1}
\end{figure}
  
\begin{figure}
\centering
\resizebox{\hsize}{!}{\includegraphics{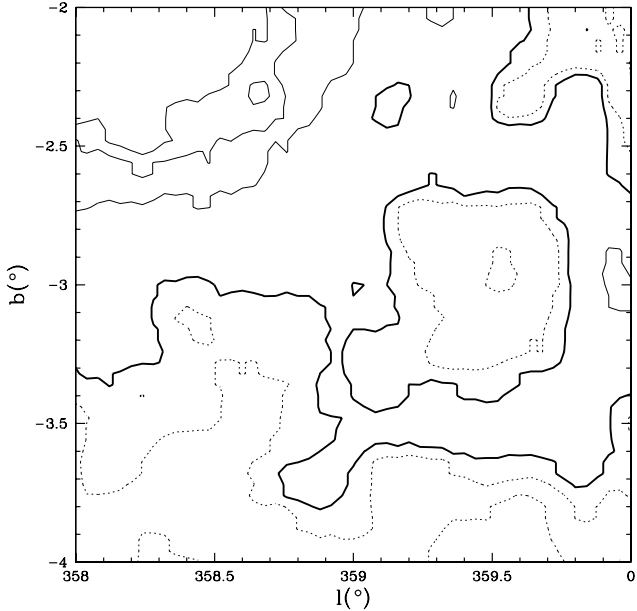}}
\caption[]{{\bf FIR extinction map of window W359.4-3.1. 
The contour levels correspond to {\it $A_{KFIR}$} = 0.31 and 0.38 (dotted lines), 0.42 (thick 
solid line), and 0.59, 0.76 and 0.87 (thin solid lines).}}
\label{fig1}
\end{figure}

\subsection {Deriving $A_K$ from 2MASS data}

{\bf An alternative way to obtain K band extinction values and to
build extinction maps for W0.2-2.1 and W359.4-3.1 is to use
the 2MASS {\it JK$_s$} photometry available in 
Web Interface {\it http://irsa.ipac.caltech.edu/applications/Gator/}. 2MASS also provides {\it H} band data, but
we prefer not to use them, since the ({\it J}-{\it K$_s$}) colours
appear to be more sensitive to extinction variations than the 
({\it H}-{\it K$_s$}) ones; this is probably due to the larger wavelength interval.} 

We extracted the 2MASS data for stars with 8.0 $\leq$ {\it K$_s$} $\leq$ 11.5 
within 1$^{\circ}$ radius centred on the Galactic coordinates 
{\bf ($\ell$,{\it b}) = (0.0$^{\circ}$, -2.0$^{\circ}$)} and {\bf ($\ell$,{\it b}) = 
(359.0$^{\circ}$,-3.0$^{\circ}$)}. 
These {\bf are the same centers as in the FIR extinction maps discussed in
the previous subsection.
The choice of {\it K$_s$} magnitude range for data extraction is motivated
by the fact that} the upper giant branch is as well defined and linear 
in this range as in {\bf Baade's window}. The total number of extracted stars 
from the 2MASS database were 90,407 in the W0.2-2.1 region and 
69,286 in the W359.4-3.1 region. For comparison, we also extracted 
2MASS data from regions at {\bf ($\ell$,{\it b}) = (1.0$^{\circ}$,-3.0$^{\circ}$)} and  
{\bf ($\ell$,{\it b}) = (1.0$^{\circ}$, -4.0$^{\circ}$)}, which include the 
Sgr I (48,040 stars) and {\bf Baade's} (20,358 stars) windows, respectively {\bf (see
Table~1)}. We note, however, that the 2MASS archive data do not 
yet provide complete coverage of these two comparison fields. 
The mean magnitude errors from the extracted 2MASS data are 
{\it $<\sigma>_J$} = 0.04$\pm$ 0.01 and 
{\it $<\sigma>_{K_S}$} = 0.04$\pm$ 0.01. These {\bf photometric errors 
bracket} 95 \% and 92 \% of all extracted stars, respectively for {\it $J$} and 
{\it $K_S$}. 

In order to map the selected low-extinction regions in {\it $A_K$}, we define 
small square {\bf cells} with  $4^{\prime}\times 4^{\prime}$. The extinction in each cell 
was determined by upper 
giant branch fitting to its observed CMD, similar to the FTK99 extinction 
determination method.  
FTK99 derived {\it $A_K$} values for their fields using the upper giant 
branch of Baade's window (Tiede et al. 1995) as reference.  
{\bf Our reference upper giant branch was defined from that of FTK99.
We proceeded as follows: we first extracted 2MASS photometry for 
seven fields from FTK99. These fields are g0-1.8a, g0-2.3a, g0-2.8a, 
g1-1.3a, g2-1.3a, g3-1.3a and g4-1.3a, where the field designation 
incorporates its position in Galactic coordinates.}  We used the {\it K} and 
{\it $K_s$} filter transmission curves given in 
Persson et al. (1998) and the extinction curve of Cardelli et al. (1989) to 
obtain the ratios ${\it \frac{A_{K_s}}{A_V}} = 0.118$,  
${\it \frac{A_{K}}{A_V}} = 0.112$ and $ {\it \frac{A_K}{A_{K_s}}} = 0.95$. 
These ratios allow us to transform from ${\it A_{K_s}}$ to {\it $A_K$}. The 
relation between 
extinction and reddening ${\it A_K} = 0.618 {\it E(J-K)}$ (Mathis 1990) 
was used to derive the relation 

\begin{equation}
A_{K_s} = 0.670~E(J-K_s). 
\end{equation}

We then extinction corrected the 2MASS CMDs of each of the 7 FTK99 fields 
and made a composite corrected CMD. Figure 4 shows the final CMD based on 
2MASS data in FTK99's fields.{\bf It is clearly dominated
by an upper giant branch, whose mean locus is shown as a straight line.}
This latter is the result of 
a linear fit to the upper giant branch in the adopted magnitude range, yielding

\begin{equation}
(K_S)_0 = -7.81*(J-K_S)_0+17.83.
\end{equation}

\begin{figure}
\centering 
\resizebox{\hsize}{!}{\includegraphics{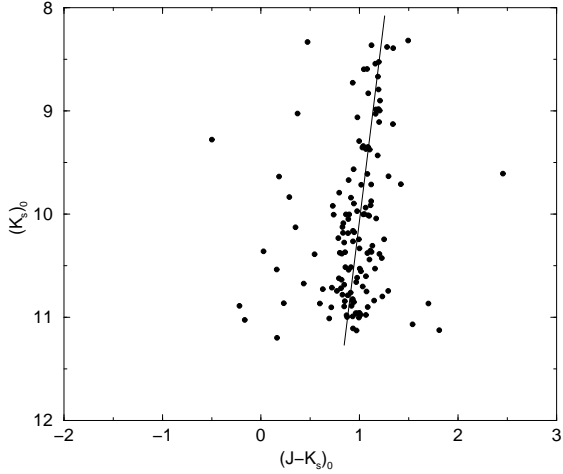}}
\caption[]{Determination of the upper giant branch mean locus from the
{\bf combined CMD of the  fields in Frogel et al. (1999).}}
\label{fig1}
\end{figure}

Assuming that the upper giant branch in each of our 
$4^{\prime}\times 4^{\prime}$ cells is
comparable to and has the same slope as the  extinction-corrected
template given by equation {\bf (2)}, we can infer the extinction 
in each cell {\bf within the W0.2-2.1 and W359.4-3.1 regions.}  
From the {\it ($K_s$,J-$K_s$)} CMD values of each star in the cell, 
we calculated the shift along the reddening vector {\bf given by equation} {\bf (1)}
required to make 
it fall onto the reference upper giant branch.  
The {\it $A_K$} value for each 
cell was taken as the median of the distribution of such values. Foreground 
contamination was minimized by applying a 2-$\sigma$ clipping to the 
data and recalculating the 
median extinction and deviation iteratively until convergence. {\bf We hereafter
use the notation $A_{K,2MASS}$ to denote the extinction value determined in
this way.}

In panels (5a) and (5c) 
we show extinction-corrected {\it (K$_s$)$_0$}, {\it (J-K$_s$)$_0$} CMDs 
for stars of typical cells in the W0.2-2.1 and W359.4-3.1 maps, 
respectively. 
Each CMD was corrected by the median extinction derived
from the histograms of {\it $A_K$} values in each cell. 
These histograms are shown
in panels (5b) and (5d) for {\bf the cells in} W0.2-2.1 and W359.4-3.1 maps, respectively,
their median values being {\it $A_{K,2MASS}$} = 0.54$\pm$ 0.16 and 
{\it $A_{K,2MASS}$} = 0.33$\pm$0.10. 
We note that there are secondary upper giant branches in the CMDs 
and consequently a second peak in {\it $A_K$} distribution for the two cells 
(at {\it $<A_K>$} = 0.14 and 
0.02 for the chosen cells in W0.2-2.1 and W359.4-3.1, respectively). 
These are probably caused 
by a secondary dust layer or to a dust cloud variation scale smaller than 
4 arcminutes along the lines of sight considered.

\begin{figure}
\centering 
\resizebox{\hsize}{!}{\includegraphics{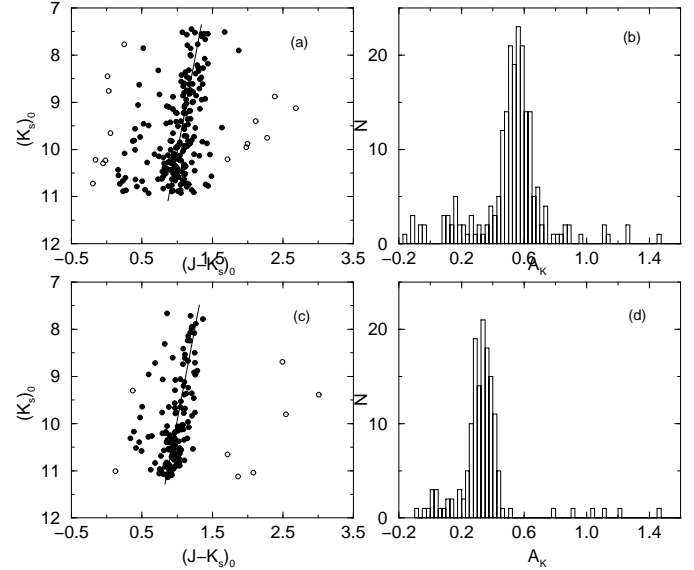}}
\caption[]{(a) {\it (K$_s$)$_0$}, {\it(J-K$_s$)$_0$} CMD for the cell at 
$\ell$ = 0.07$^{\circ}$ {\it b} = -1.2$^{\circ}$ in the W0.2-2.1 field; 
(b) {\it $A_K$} histogram derived from CMD in (a) as described 
in the text; (c) {\it (K$_s$)$_0$}, {\it (J-K$_s$)$_0$} CMD for the cell at 
$\ell$ = 359.2$^{\circ}$  {\it b} = -2.2$^{\circ}$ in the W359.4-3.1 field; 
(d) {\it $A_K$} histogram derived from CMD in (c) as described 
in text. The open circles in the CMDs indicate stars rejected due to the 
2-$\sigma$ clipping of 
the {\it $A_K$} distribution. The solid lines in the CMDs indicate the 
reference 
upper giant branch locus (Eq. 2).}
\label{fig1}
\end{figure}

For the comparison regions, which include the
low-extinction Sgr I and Baade's windows, we 
obtain {\it $A_{K,2MASS}$} = 0.23$\pm$0.05 and {\it$A_{K,2MASS}$} = 
0.18$\pm$0.04, respectively, using the same method described above. 
Glass et al. (1995) studied variable stars 
in the Sgr I window and adopted {\it $A_K$}=0.21. For Baade's Window, 
considering the extinction map obtained by Stanek (1996) and zero-point 
calibrations (Gould et al. 1998 and Alcock et al. 1998), the mean extinction 
is {\it $<A_K>$} = 0.17$\pm$0.03. {\bf The adopted extinction determination method and the 2MASS photometry produce 
results compatible to those found in the literature.} Thus, they are certainly 
useful to map the extinction in the inner Galactic Bulge, although one 
has to keep in mind that 
regions much closer to the Galactic Centre are more affected by crowding 
and highly variable 
extinction, which may in turn have an impact on photometric precision and 
on the applicability of the method.
Another issue is the possible existence of metallicity gradients in the stellar
content of the inner Bulge, which may result in systematic errors in 
the inferred extinction values (Schultheis et al. 1999). However, there 
is some recent evidence,
based on spectroscopy of M giants, against significant metallicity variations
in that region (Ram\'\i rez et al. 2000).

\begin{figure}
\centering 
\resizebox{\hsize}{!}{\includegraphics{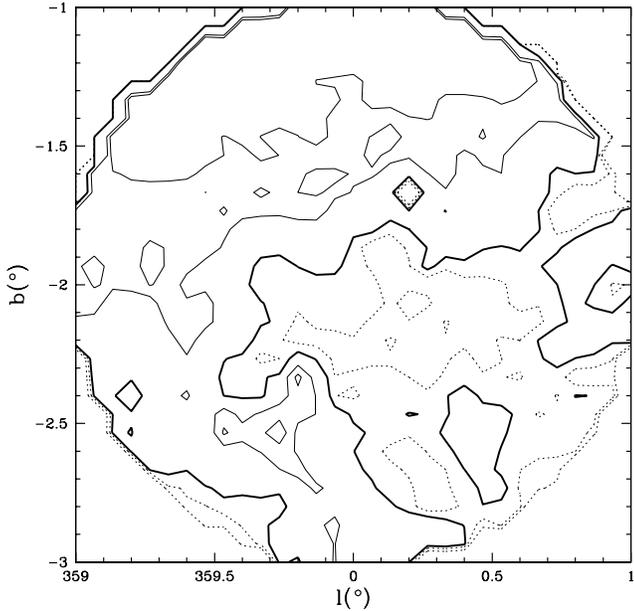}}
\caption[]{{\bf 2MASS extinction map  
of window W0.2-2.1. Contour 
levels correspond to {\it $A_{K,2MASS}$} =  0.2 and 0.25 (dotted lines), 
0.29 (thick solid 
line), and 0.4 and 0.5 (thin solid lines).}}
\label{fig1}
\end{figure}

\begin{figure} 
\resizebox{\hsize}{!}{\includegraphics{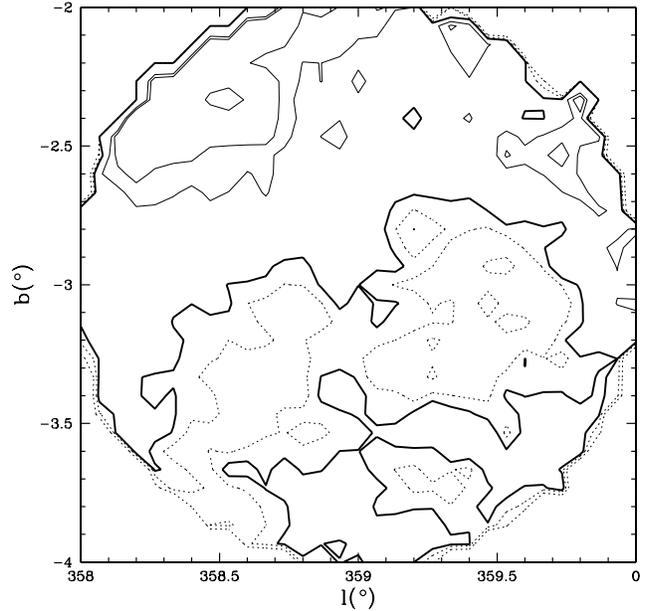}}
\caption[]{{\bf 2MASS extinction map of window  
W359.4-3.1. Contour 
levels correspond to {\it $A_{K,2MASS}$} = 0.2 and 0.25 (dotted lines), 
0.28 (thick solid 
lines), and 0.4 and 0.5 (thin solid lines).}}
\label{fig1}
\end{figure}

\subsection {2MASS extinction maps}

Figures 6 and 7 show the {\it $A_{K,2MASS}$} maps for the W0.2-2.1 and 
W359.4-3.1 regions, respectively. {\bf These contour maps are similar in shape
to those shown in Figures 2 and 3. In both figures we see nearly
closed regions with $0.2 < A_{K,2MASS} < 0.25$.
The mean extinction values
are {\it $<A_{K,2MASS}$} = 0.29 $\pm$ 0.05 in W0.2-2.1 and 
{\it $<A_{K,2MASS}>$} = 0.28 $\pm$ 0.04 in W359.4-3.1. Therefore, some cells
within the low-extinction windows have
$A_{K,2MASS}$ value about 2 std below the mean map values.
Note that the lack of contours
in the corners of the figures just reflects the circular areas used for
the 2MASS extractions.}

Panels (a) and (c) of Figure 8 show the histograms 
of {\it $A_{K,2MASS}$} values for all
the cells in the W0.2-2.1 and W359.4-3.1 maps. 
In W0.2-2.1 there is a larger number
of cells with high {\it $A_{K,2MASS}$} values than in W359.4-3.1. 
This {\bf reflects the closer proximity of the former field to 
the Galactic Centre.} 
Panels (b) and (d) of the same figure show 
the internal errors in 
the determination of the {\it $A_{K,2MASS}$} values as a function 
of {\it $A_{K,2MASS}$} itself.
These errors correspond to the std of the final, sigma-clipped,
histogram of {\it $A_{K,2MASS}$} values in each cell.
We notice that in W0.2-2.1 the internal errors have a larger
dispersion with {\it $A_{K,2MASS}$}. Such behaviour was also found
by FTK99 and is due to the small scale variations of
the dust distribution in the region. The mean internal errors 
are $<\sigma>$ = 0.09 $\pm$ 0.03
in W0.2-2.1 and $<\sigma>$ = 0.08 $\pm$ 0.03 in W359.4-3.1. 
These errors are somewhat higher than
those estimated by FTK99 in their Bulge fields. The reason is that
that our 4$^{\prime} \times$ 4$^{\prime}$ cells
are larger than those used by FTK99 (1.5$^{\prime} \times$ 1.5$^{\prime}$) 
and should, therefore, include a larger dispersion by dust gradients.
  
The bulge windows W0.2-2.1 and W359.4-3.1 are relatively closer to the 
Galactic Centre than Sgr I, Sgr II and Baade's windows 
and are located in a hole surrounded by the dark clouds LDN48, 
LND43, LDN1801, LDN1769,  LDN1783, LDN3, LDN1795, LDN1788 from the  Lynds' 
(1962) catalogue and 
FSDN435, FSDN431, FSDN430, FSDN444 from the Feitzinger \& St\"uwe's 
(1984) catalogue.

\begin{figure} 
\resizebox{\hsize}{!}{\includegraphics{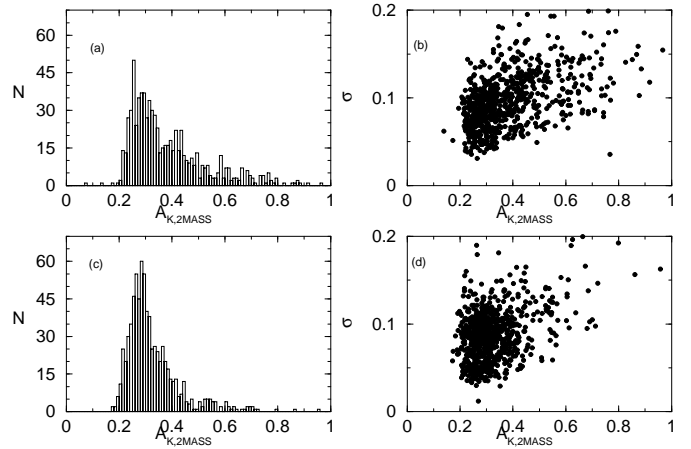}}
\caption[]{{\bf 2MASS extinction histogram for window
 W0.2-2.1  in Panel (a),  and for W359.4-3.1 in Panel (c). 
 Extinction internal errors as a function of {\it $A_{K,2MASS}$}  for 
W0.2-2.1  in Panel (b), and for W359.4-3.1 in Panel (d).}}
\label{fig1}
\end{figure}

\subsection{Comparison between 2MASS and FIR emission extinction maps}

\begin{figure}
\centering 
\resizebox{\hsize}{!}{\includegraphics{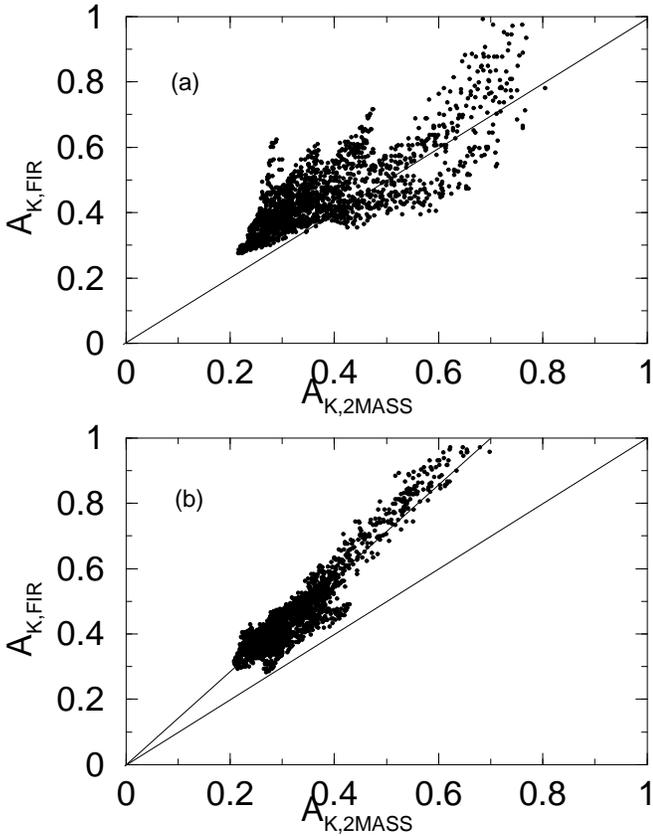}}
\caption[]{{\bf Comparison of FIR and 2MASS extinction values  for cells in 
windows W0.2-2.1 (panel (a)) and  W359.4-3.1 (panel (b)). The identity function is indicated as 
a straight line. The function ${\it A_{K,FIR}}=1.45 {\it A_{K,2MASS}}$ is also indicated in panel (b).}}
\label{fig1}
\end{figure}

{\bf As mentioned previously, the 2MASS extinction maps of Figures 6 and 7
show essentially the same features as the FIR extinction maps 
(Figures 2 and 3).
In order to investigate this similarity in a more quantitative manner, 
we convolve the 2MASS {\it $A_K$} extinction maps with 
a $\sigma$ = 4.5$^{\prime}$ Gaussian, obtaining a resolution 
of approximately 6$^{\prime}$ FWHM, which
is compatible with that of SFD98's extinction maps (6.1$^{\prime}$). 
Figure 9 shows the comparison between the $A_{K,2MASS}$ and
$A_{K,FIR}$ values for both low-extinction windows.

In both panels we see a clear correlation between 
{\it $A_{K,FIR}$} and {\it $A_{K,2MASS}$}, 
quantitatively confirming the similarities between the two extinction maps.
The relation between the two $A_K$ estimates, however, has a much more complex 
pattern in W0.2-2.1 (panel 9a) than in W359.4-3.1 (panel 9b).} We see several branches
in panel {\bf (9a)}, some above the identity line, some lying below it. 
We verify that the cells that form these
particular structures in the {\it $A_K$} scatter plot correspond to 
specific spatial regions in the extinction map. 
These may correspond to lines of sight crossing specific dust clouds whose
physical characteristics, most specially dust temperature and density, could
yield distinct signatures in their emission and absorption properties.
As mentioned in the end of the previous section, several individual
dark clouds lying close to and around our two low-extinction regions have
been catalogued. 

In the W359.4-3.1 region (panel 9b), there is a {\bf distribution much closer to 
linear relation between $A_{K,2MASS}$ and $A_{K,FIR}$.}
The slope in this relation, however, 
is larger than unity: ${\it A_{K,FIR}}=1.45 {\it A_{K,2MASS}}$.
Arce \& Goodman (1999) obtained a similar result in their analysis of the
Taurus dark cloud complex ({\it b}$\approx$ -15$^{\circ}$); 
by comparing the dust emission values derived by SFD98 
to those derived from four other methods,
including a dust emission extinction derived directly from 100 $\mu$m 
flux with temperature corrections.
They concluded that SFD98 may be overestimating extinction
by a factor varying from 1.3 to 1.5.
{\bf They also pointed out that this factor could not be 
due to overestimation of the ratio of total-to-selective extinction {\it $R_V$}, 
because several studies (Kenyon et al. 1994; Vrba \& Rydgren 1985; and recently Whittet 
et al. 2001) suggested that the  {\it $R_V$} = 3.1 in most parts of the Taurus 
dark cloud complex, in regions where the visual extinction is lower than {\it $A_V <$} 
3.0.
This systematic effect in $A_{K,FIR}$ may originate
from SFD98's calibration of the dust 
column density $\times$ reddening relation, which was derived 
in regions of low to moderate extinction ({\it E(B-V)}$\simeq 0.15$), but
may not apply equally well to the high extinction regime.

As a helping tool in understanding} the differences in 
the {\it $A_{K,FIR}$} and 
{\it $A_{K,2MASS}$} values, we have considered the amount of dust expected to
lie beyond the inner Galactic Bulge, therefore being located 
behind most of the stars measured by 2MASS. This dust located on the far 
side of the Galaxy will have an obvious influence
over the {\it $A_{K,FIR}$} $\times$ {\it $A_{K,2MASS}$} correlation, 
since it is expected to
affect the first but not the second extinction value. 
A simple model of dust distributed on
a plane with an exponential drop both along and perpendicular
to this plane has been used to predict the relative contribution 
of the material
on the far side of the Galaxy. We thus consider a model of linear absorption
coefficient given by:

$$\sigma_{\lambda} \propto e^{-R/R_d} e^{-Z/Z_d},$$

where {\it $R$} and {\it $Z$} are cylindrical coordinates centred on the 
Galaxy, whereas
{\it $R_d$} and {\it $Z_d$} are the dust horizontal and vertical 
exponential scales, respectively. By adopting reasonable values for
the scale parameters (such as ${\it R_d} \simeq 2.5~kpc$ and 
${\it Z_d} \simeq 100~pc$)
and integrating the dust absorption along the two lines
of sight where our low-extinction windows are located, we conclude that 
$ < 20\%$ of the total extinction towards W0.2-2.1 is caused by dust on 
the background of the stars belonging to the inner Bulge in this direction.
For W359.4-3.1, this amount is 5\% or less. Therefore, we conclude that
dust on the far side of the Galaxy may partially account for 
the differences in the DIRBE/IRAS and 2MASS {\it $A_K$} determinations, 
especially for {\it $A_{K,FIR}$} being larger 
than {\it $A_{K,2MASS}$} in most cells. On the other hand, 
background dust does not explain the complex 
patterns seen in W0.2-2.1 or the slope in the W359.4-3.1 scatter plot.

{\bf Temperature effects may play a role in explaining the 
observed features, since dust clouds in the direction of the Galactic Centre may be warmer on average than
elsewhere.} In addition, the low resolution ($\approx$ 1$^{\circ}$) 
of SFD98's temperature maps {\bf increases the uncertainties in the 
$A_{K,FIR}$ temperature corrections, especially in zones of
large temperature gradients.
Figure 10 shows SFD98's temperature map in the central 10$^{\circ}$ of the Galaxy. Despite the low
resolution, we notice that W0.2-2.1 is much closer to a local peak
in the dust temperature map than W359.4-3.1.
This adds support to the idea of temperature gradients 
contributing to the complexity of panel (9a) as compared to panel (9b).}

{\bf Additional sources of scatter in the $A_{K,2MASS} \times A_{K,FIR}$
relation are possible contamination effects in both datasets. 
As pointed out by SFD98, extragalactic and unresolved Galactic 
sources at low Galactic latitudes (${\it |b|} <$ 5$^{\circ}$) have not been removed from the dust maps.
 They could be an important
contribution to increase {\it $A_{K,FIR}$} values in very low latitudes, 
in particular in the
Bulge window closer the Galactic Plane, W0.2-2.1. Concerning the 2MASS data, apart
from applying the $\sigma$ clipping of the $A_K$ distribution in each cell,
we did not attempt any further correction for contaminating foreground
disk stars.}

\begin{figure}
\centering 
\resizebox{\hsize}{!}{\includegraphics{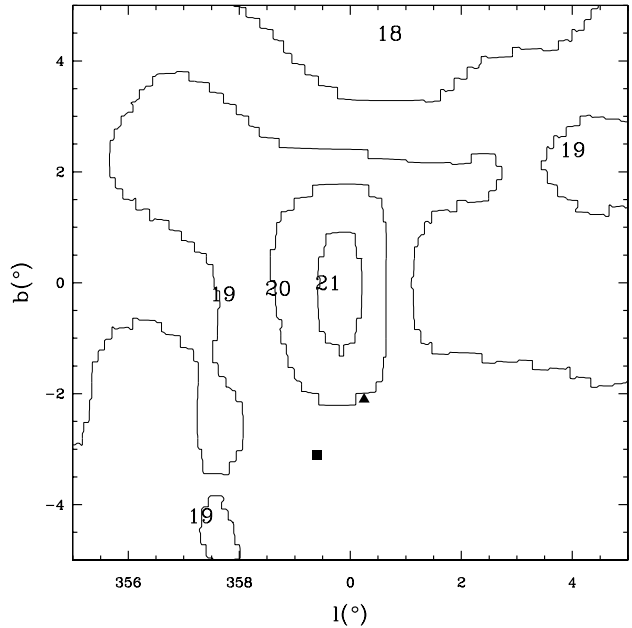}}
\caption[]{{\bf 10$^{\circ}\times 10^{\circ}$  temperature map centred on the 
Galactic Nucleus,
according to Schlegel et al. (1998). The temperature contours are
given in Kelvin. The triangle and square correspond to the position of 
W0.2-2.1 and W359.4-3.1, respectively.}}
\label{fig1}
\end{figure}

\section{Concluding remarks}

We identified a new low-extinction window towards the Galactic Bulge, 
W359.4-3.1, located at {\bf ($\ell$,{\it b}) = (359.40$^\circ$,-3.10$^\circ$) and 
of size} $40^{\prime}\times 30^{\prime}$, using a 
10$^{\circ}$ $\times$ 10$^{\circ}$ map around the Galactic Centre 
extracted from Schlegel et al.'s (1998) FIR emission reddening database.

Using the infrared photometry available from the 2MASS {\it JHK$_s$} survey 
archive, we built {\it $A_{K,2MASS}$} extinction maps for 
W359.4-3.1 and for W0.2-2.1, this latter 
window previously identified by Stanek (1998). The extinction values were 
determined by means of upper giant branch fitting, using as a reference 
the upper giant branch of fields previously studied by 
Frogel et al. (1999). The extinction determination method and the 2MASS 
photometry were tested on fields with well known low-extinction values: 
Sgr I and Baade's Window (Baade 1963). Our 
derived extinction values for these windows agree well with previous values 
quoted in the literature, {\bf indicating that the present method is a useful tool for the study of the bulge windows}.

The {\it $A_{K,2MASS}$} maps confirmed the existence of the two bulge windows. 
The mean extinction in the field around W0.2-2.1 is 
{\it $<A_{K,2MASS}>$} = 0.29 
$\pm$ 0.05, whereas in the field around W359.4-3.1 we obtained 
{\it $<A_{K,2MASS}>$} = 0.28 $\pm$ 0.04. 
In both cases we find {\bf an area of systematically lower than
average {\it $A_{K,2MASS}$} values, with minima around $A_{K,2MASS} \simeq 
0.20$.} These windows are located very close to the Galactic Centre, through a 
hole in the distribution of known dark clouds. 
{\bf The extinction maps} obtained with 2MASS data show very similar features 
to those {\bf based on FIR dust emission data}. 
However, some systematic effects, usually
in the sense that ${\it A_{K,FIR}} > {\it A_{K,2MASS}}$, are seen 
when a direct comparison of the two extinction estimates is made. 
In particular, for the W359.4-3.1 region, ${\it A_{K,FIR}} \simeq 1.45 \times {\it A_{K,2MASS}}$ with a small scatter. For W0.2-2.1, which lies closer to
the Galactic Centre, the situation is more complex, with several structures
with distinct correlations between dust emission and absorption being present.

The qualitative agreement suggests that the dust clouds that 
redden the bulk of Bulge stars 
are the main contributors to the dust emission in the region. Indeed, 
a simple model for the distribution of dust shows that 
most of the dust clouds 
in the line of sight towards the low-extinction windows should be located 
{\bf on the near side of the Galaxy}. The amount of dust expected to lie
beyond {\bf the Galactic Centre} does not alone explain the quantitative 
differences between
{\it$A_{K,FIR}$} and {\it$A_{K,2MASS}$}. These may be caused by temperature
variations as a function of distance from the centre of the Galaxy or
from one individual dust cloud to another. {\bf Alternatively, systematic effects in the 
conversion of DIRBE/IRAS maps into extinction measures might explain  the 
observed differences. It is not possible} to
disentangle these effects using only the data shown here.

These low-extinction windows are interesting new targets to studies 
of the properties of the Bulge stellar population, {\bf especially considering
their location closer to the Galactic Centre than Baade's Window.}

\begin{acknowledgements}
We thank Dr. D. Schlegel for pointing out to us the availability
of a new version of his software {\it dust-getval} for reading the temperature
maps.
This publication makes use of data products from the Two Micron All 
Sky Survey, which is a joint 
project of the University of Massachusetts and the Infrared Processing 
and Analysis Center/California Institute of Technology, funded by the National Aeronautics 
and Space Administration and the National Science Foundation. {\bf We also have made use of the NASA/ IPAC Infrared Science Archive,
 which is operated by the Jet Propulsion Laboratory, California Institute of Technology, under contract with the NASA. We acknowledge support from the 
Brazilian institutions FAPESP and CNPq. CD acknowledges the FAPESP pos-doc fellowship
proc. 00/11864-6.} 
\end{acknowledgements}

\end{document}